\documentclass[sigconf]{acmart}

\AtBeginDocument{%
  \providecommand\BibTeX{{%
    \normalfont B\kern-0.5em{\scshape i\kern-0.25em b}\kern-0.8em\TeX}}}

\copyrightyear{2026}
\acmYear{2026}
\setcopyright{cc}
\setcctype{by-nc-nd}
\acmConference[DIS '26]{Designing Interactive Systems Conference}{June 13--17, 2026}{Singapore, Singapore}
\acmBooktitle{Designing Interactive Systems Conference (DIS '26), June 13--17, 2026, Singapore, Singapore}
\acmDOI{10.1145/3800645.3812902}
\acmISBN{979-8-4007-2563-0/2026/06}

\usepackage{enumitem}
\usepackage{colortbl}
\usepackage{multirow}

\usepackage{graphicx}
\usepackage{float}

\definecolor{contextbg}{RGB}{230,240,248}   
\definecolor{conflictbg}{RGB}{248,235,235}  
\definecolor{managebg}{RGB}{249,246,230}    

\begin{document}

\title[]{Designing with Tensions: Older Adults' Emotional Support-Seeking Under System-Level Constraints in Conversational AI}



\author{Mengqi Shi}
\affiliation{
  \institution{University of Washington}
  \city{Seattle}
  \country{United States}
}
\email{shi21@uw.edu}

\author{Tianqi Song}
\affiliation{
  \institution{National University of Singapore}
  \city{Singapore}
  \country{Singapore}
  }
\email{tianqi_song@u.nus.edu}

\author{Zicheng Zhu}
\affiliation{
  \institution{National University of Singapore}
  \city{Singapore}
  \country{Singapore}
  }
\email{zicheng@u.nus.edu}

\author{Yi-Chieh Lee}
\affiliation{
  \institution{National University of Singapore}
  \city{Singapore}
  \country{Singapore}
}
\email{yclee@nus.edu.sg}

\begin{abstract}
  Older adults have increasingly turned to conversational AI as a source of emotional support. However, little is known about how emotionally supportive interactions are experienced in everyday use, particularly when AI systems limit, redirect, or intervene during these interactions. We interviewed 18 older adults about their experiences using conversational AI for emotional support, examining when they turn to AI, how they engage during emotionally vulnerable moments, and how they respond when support feels disrupted. 
Our findings show that older adults often rely on AI when other forms of social support feel inaccessible. However, current safety-related interventions can redirect interactions in ways that participants experience as interruptions to emotional engagement or as shifts in control away from them. Such disruptions can undermine older adults’ ability to remain emotionally engaged and, in some cases, contribute to emotional distress.
We discussed design implications for emotionally supportive conversational AI, emphasizing the need for safety interventions that are enacted within older adults' social contexts, align with users' emotional pacing, and preserve their sense of agency.
\end{abstract}




\maketitle


\section{Introduction}

For many older adults, later life involves gradual changes in health \cite{ouden2021impact}, social networks \cite{vos2020exploring, rook2017close}, and daily routines \cite{o2019daily}, which can increase the need for emotionally responsive support. Emotional concerns are not always easily shared through existing social relationships, even when such relationships are present \cite{ong2016loneliness, ho2023examining, palmer2016does}. These circumstances can reduce opportunities for informal emotional expression and limit available alternatives when emotional support is needed \cite{wang2024techroles, hornung2017ICT}, and they have been associated with increased risks of loneliness \cite{berridge2023companion}. As a result, companionship technologies have attracted growing interest in later life \cite{jones2021reducing, zhang2025rise}. Within this context, many older adults express strong expectations for empathy and responsiveness in conversational AI, particularly during emotionally vulnerable moments \cite{rodriguez2023qualitative, huang2025review, razavi2022discourse}. In recent years, these expectations have increasingly been directed toward conversational and AI-based technologies that can offer readily available forms of interaction and support, positioning AI-based companionship and emotionally supportive technologies as meaningful resources in later life.

Building on these expectations, conversational AI systems have become increasingly embedded in everyday routines \cite{liu2024understanding, yang2019understanding}. Beyond task-oriented assistance, contemporary conversational systems support open-ended dialogs and simulate empathic responses, extending their use into emotionally intimate domains \cite{ruane2019consideration, irfan2024recommendations}. Accordingly, conversational AI is increasingly framed and explored as a scalable form of support in contexts involving loneliness, emotional distress, and limited access to consistent human support \cite{wang2024techroles, alnefaie2021usageoverview}. Public discussions and media coverage have also described instances in which older adults use AI chatbots for companionship when human support is unavailable \cite{forbes2025lonely, cbs2025lonely}.

As emotionally supportive uses of conversational AI expand, concerns have also emerged around potential harms associated with these interactions. Prior work has highlighted a range of issues related to vulnerability, reliance, and trust in emotionally supportive AI, alongside reported benefits such as reduced loneliness \cite{xie2024longitudinal, bates2024snugglebot, aikaterina2025characterizing}. Researchers have also drawn attention to concerns about relational transgressions, privacy violations, misinformation, and risks related to harmful dependency or inappropriate emotional influence \cite{zhang2025dark, saxena2025potentialharms, shelby2023sociotechnical}. These issues have become increasingly salient in discussions of AI companionship, particularly in relation to emotionally intimate and supportive use contexts \cite{jones2025personalization}.

Prior research has primarily addressed AI safety in conversational systems through system-level interventions such as content filtering \cite{khalatbari2023learnlearngenerativesafety}, redirection \cite{meade2023usingincontextlearningimprove}, or conversation termination \cite{zheng2025leteasycontextualeffects} once predefined risk thresholds are reached. Much of this work focuses on how safety responses, such as refusals or redirections, can be phrased or softened to reduce immediate user harm. Similarly, industry alignment practices have emphasized limiting over-anthropomorphism and emotionally manipulative interactions in deployed large language model (LLMs) \cite{jones2025personalization, manzini2025bindsus, zhang2025dark}. Across both research and industry contexts, these approaches are commonly grounded in the assumption that interrupting or constraining interactions increases user safety.

As emotionally supportive AI becomes more common in older adults’ everyday lives~\cite{tang2025ai, song2026understanding}, questions emerge around how these interactions are experienced when safety interventions shape or interrupt ongoing support. For many older adults, emotionally supportive interaction is expected to allow continued engagement, emotional expression on their own terms, and influence over how and when an interaction progresses. These expectations are shaped by social contexts marked by reduced social networks and a preference for emotionally meaningful interactions over time \cite{carstensen1999taking}, as well as strong values placed on autonomy and control in relationships and decision-making \cite{taylor2024older}. When safety mechanisms designed for broad user populations intervene during emotionally vulnerable moments, older adults may experience these interventions as abrupt interruptions, rejection, or a loss of control rather than as support. Such experiences may lead to unintended emotional harm or disengagement.

These limitations point to a clear research gap. Much existing work on AI safety and emotional support focuses on system-level risks \cite{raman2025intolerable} or relies on social media analysis \cite{chu2025illusions} and conversation-level modeling \cite{liu2021towards}, with limited attention to older adults’ lived experiences during emotionally supportive interactions. What remains unclear is how older adults experience, navigate, and respond to safety interventions during emotionally supportive interactions in everyday use, and how these experiences should inform design decisions about balancing safety with users’ needs in emotionally supportive contexts.

To address this gap, we examine older adults’ experiences with emotionally supportive AI in everyday use, focusing on the contexts in which older adults turn to AI for emotional support, how emotionally supportive interactions are shaped by safety mechanisms, and the practices older adults use to maintain a sense of agency and meet emotional needs. Rather than treating safety as a system-level outcome enforced through constraint, we focus on how emotionally supportive interactions are experienced over time and how safety-related boundaries and user agency shape these experiences.

This study explores three key research questions:
\begin{itemize}[topsep=0pt, itemsep=1pt]
    \item \textbf{RQ1}: How and under what circumstances do older adults turn to conversational AI for emotional support in everyday life?
    \item \textbf{RQ2}: What tensions and breakdowns do older adults experience during emotionally supportive interactions with AI?
    \item \textbf{RQ3}: How do older adults respond to and navigate safety interventions during emotionally supportive interactions with AI, and what unmet needs or expectations do these responses reveal for emotionally supportive AI systems?
\end{itemize}

Our study makes three primary contributions. First, we provided empirical evidence on how older adults use conversational AI for emotional support and navigate emotionally vulnerable moments in everyday life. Drawing on older adults' first-hand accounts, we show that age-related changes in social availability and the motivational shifts described by Socioemotional Selectivity Theory \cite{carstensen1999taking} shape both why older adults turn to AI for support and why disruptions to these interactions carry particular weight in later life.

Second, we identified an interactional mismatch between system-level safety responses and process-oriented emotional support, demonstrating how safety interventions can unintentionally disrupt emotional continuity and undermine users' perceived sense of agency during support-seeking interactions. For older adults, whose reliance on AI as a primary support channel is more likely to be structural \cite{carstensen1999taking, tang2020systematic} rather than situational, the interactional costs of safety interventions become harder to absorb. This mismatch carries interactional costs that age-neutral safety designs are not equipped to address.

Finally, we derived interaction-centered design implications that reconceptualize safety not merely as a system constraint but as a design responsibility that must attend to timing, continuity, and users' lived experiences within ongoing emotional interactions.

\section{Related Work}

\subsection{AI Companionship for Older Adults}

As global aging accelerates \cite{hernandez2025exposome}, increasing attention has been paid to how artificial intelligence can support older adults who live alone or spend long periods at home, particularly in emotional and social domains \cite{wang2024techroles, syed2024roleofai}. Early work on companionship technologies primarily focused on physical companion robots, which aimed to reduce loneliness and provide psychological comfort through continuous presence and interaction \cite{jeong2023roboticcompanion, manzini2025bindsus, mun2025particip-ai, halln2001slow}. Prior studies describe how such systems can create a sense of being accompanied or listened to in everyday life \cite{berridge2023companion, jones2021reducing}.

With the widespread adoption of voice assistants, older adults no longer rely solely on specially designed robots to access companion-like interactions \cite{upadhyay2023long-term, even2022benefits}. Research shows that many older adults treat voice assistants as entities they can talk to at any time, valuing responsiveness, patience, and non-judgmental language in emotionally oriented interactions \cite{trajkova2020reasonsforusing, even2022benefits, larubbio2025navigating, cuadra2023designing}.

More recently, the emergence of LLMs has expanded the scope of AI companionship \cite{jones2025personalization, chen2023intimate}. Compared to earlier systems, LLM-generated language supports longer and more expressive conversations \cite{zhang2025rise}, encouraging self-disclosure, personal storytelling, and perceptions of emotional understanding \cite{enam2025percptions, shahid2025personalit, guo2025selfdisclosure}. Some studies report that older adults experience heightened social presence in these interactions and may treat conversational AI as partners comparable to familiar social contacts \cite{chin2024styles, jones2021reducing, guo2025selfdisclosure}. In specific contexts, such as early-stage cognitive impairment, conversational AI has also been explored as a resource for everyday social engagement \cite{xygkou2024mindtalker, jeong2023roboticcompanion}.

Across this body of work, AI companionship for older adults has evolved from embodied robotic systems to voice-based and general-purpose conversational technologies \cite{ventura2025ageofai, huang2025review, mun2025particip-ai, song2026understanding}. Existing studies primarily document patterns of adoption, interaction, and perceived companionship in emotionally supportive use. What remains less examined, however, is how older adults experience these interactions when system-level constraints enter ongoing emotional engagement. Given that socioemotional selectivity in later life orients older adults toward emotionally meaningful exchanges \cite{carstensen1999taking}, understanding how safety mechanisms shape the quality of such interactions warrants closer attention.

\subsection{Safety and Interactional Concerns in Human-AI Interaction}

As interest in AI companionship grows, research has increasingly examined the risks and harms associated with emotionally oriented AI systems, extending beyond a primary focus on benefits \cite{chandra2025psychologicalrisks, zhang2025dark, rheu2024chatbot, shelby2023sociotechnical}. A substantial body of work identifies algorithmic harms related to privacy, identity, and discrimination \cite{abercrombie2024taxnomy, shelby2023sociotechnical, blodgett2022responsible}, manipulation, and mental health issues \cite{zhang2025dark, abercrombie2024taxnomy}. In conversational AI, these concerns have been examined in interactional contexts, highlighting how system behaviors, user characteristics, and psychological outcomes intersect during situated use \cite{chandra2025psychologicalrisks, tahaei2023human-centered, chivukula2023wrangling}.

Within emotional and relational domains, prior studies describe how empathetic language, emotional mirroring, and persistent availability can foster strong attachments and para-social bonds between users and AI systems \cite{jones2025personalization, chu2025illusions, ventura2025ageofai}. Large-scale analyzes show that conversational agents frequently provide affirmation and emotional alignment in sensitive situations involving aggression or self-harm, illustrating ongoing tensions between responsiveness and safety-oriented constraints 
\cite{chu2025illusions, chandra2025psychologicalrisks}.

For older adults, interactional safety concerns become particularly salient in emotionally supportive AI use, as these interactions often unfold alongside age-related considerations in technology use, dependency, and boundary negotiation \cite{even2022benefits, huang2025review, glazko2025generative, zou2024mitigating}. Because emotionally oriented conversations frequently involve personal disclosure and sustained engagement, prior work reports that older adults adopt strategies such as limiting disclosure, testing system responses, or maintaining cautious trust when engaging with AI systems \cite{knowles2018distrust, murthy2021homerobot, larubbio2025navigating}. 

In care and communication contexts, where support relationships are shaped by power asymmetries and expectations of assistance, studies also document concerns related to overreach, loss of dignity, and perceived pressure to conform to system expectations \cite{hornung2017ICT, zhang2024s, saxena2025potentialharms, tahaei2023human-centered}. Prior work thus consistently attends to issues of autonomy, dignity, and boundary maintenance in later-life interactions with supportive technologies \cite{caldeira2017senior}. Yet how these concerns manifest specifically when safety mechanisms intervene during emotionally supportive AI interactions remains underexplored.

\subsection{Psychological Needs and Emotional Well-Being in Later Life}

Research in psychology and gerontology has long emphasized that emotional experiences and well-being are shaped by individuals’ ability to maintain autonomy, effectiveness, and meaningful social connections. Across contexts, these factors have been shown to influence emotional regulation, relationship quality, and psychological health. Within this broader literature, Self-Determination Theory (SDT) provides an account of how the satisfaction of basic psychological needs supports motivation and well-being \cite{deci2012self}. Its core component, Basic Psychological Needs Theory (BPNT), identifies autonomy, competence, and relatedness as central conditions for well-being across cultures and life stages \cite{vansteenkiste2020basic, tang2021contribution}.

In emotional support contexts, interactions that preserve a sense of choice, support coping capacity, and convey care are associated with more positive emotional outcomes and reduced distress \cite{shin2022social, pan2025reciprocity}. In contrast, support processes that constrain autonomy, undermine competence, or lack emotional acknowledgment have been linked to heightened stress, shame, or dependency, shaping how emotionally secure an interaction is perceived to be \cite{veale2022emotionalsafety, chandra2025psychologicalrisks}. Among older adults, BPNT has been used to explain patterns of well-being and social participation, with research suggesting increased sensitivity to autonomy loss and greater reliance on emotionally meaningful connections under aging-related constraints \cite{tang2020systematic, ferrand2014psychological, caorong2021self}. Related work in gerontology similarly highlights the importance of dignity and self-determination as functional independence declines \cite{jacelon2004concept, collopy1988autonomy, baltes2019dynamics}.

Complementing need-based accounts, Socioemotional Selectivity Theory (SST) suggests that as people age and perceive time as more limited, they increasingly prioritize emotionally meaningful goals and relationships over information seeking or novelty. This motivational shift has been used to explain older adults’ preference for emotionally satisfying interactions and their sensitivity to how social exchanges are experienced.

In human–AI interaction research, language style, response patterns, and perceived role positioning have been shown to influence whether users feel understood, able to express their needs, and able to maintain a sense of autonomy during emotionally supportive conversations \cite{chin2024styles, pan2025reciprocity, liu2021towards}.
Existing studies discuss these interactional qualities in relation to users’ psychological needs and motivational priorities, particularly in shaping how emotionally supportive interactions with AI are experienced.

\section{Methods}

\subsection{Participants: Inclusion Criteria and Recruitment}
Our study focused on older adults aged 50 and above. This age threshold was chosen to reflect prior research definitions of older adulthood in studies of technology use and emotional well-being \cite{lee2021association, juster1995overview, nakamura2022associations, wright2017psychological}. To ensure that participants could meaningfully reflect on experiences with AI-based emotional expression, we applied the following inclusion criteria: participants were required to be (1) comfortable communicating in English, and (2) have used conversational AI as a companion to express or reflect on their emotions. 

We recruited participants through a local community organization serving older adults. Recruitment materials, including a study poster and a brief description of the research, were shared with the organization, which then distributed the information via email to members of its community. Interested individuals registered by completing an online screening questionnaire. Only respondents who met all inclusion criteria were invited to participate in the study. After screening, the research team contacted eligible participants directly using the contact details they voluntarily provided.

\begin{table*}
    \centering
    \small
    \renewcommand{\arraystretch}{1.6} 
    \begin{tabular}{
        >{\raggedright\arraybackslash}p{0.8cm}
        >{\raggedright\arraybackslash}p{1.2cm}
        >{\raggedright\arraybackslash}p{1.2cm}
        >{\raggedright\arraybackslash}p{2.2cm}
        >{\raggedright\arraybackslash}p{2.2cm}
        >{\raggedright\arraybackslash}p{3.0cm}
    }
    \toprule
    \textbf{ID} & \textbf{Gender} & \textbf{Age} & \textbf{Education} & \textbf{Living Situation} & \textbf{Self-Reported AI Use Frequency} \\
    \midrule
    P1  & Male   & 76 & Associate's    & With family & Several times per month \\
    \rowcolor{gray!10}
    P2  & Male   & 52 & Master's       & With family & Several times per week \\
    P3  & Female & 77 & Associate's    & Alone       & Several times per week \\
    \rowcolor{gray!10}
    P4  & Female & 59 & Bachelor's     & With family & Several times per week \\
    P5  & Female & 66 & Bachelor's     & With family & Several times per month \\
    \rowcolor{gray!10}
    P6  & Female & 63 & Middle School  & Alone       & Several times per week \\
    P7  & Male   & 62 & Primary School & Alone       & Several times per week \\
    \rowcolor{gray!10}
    P8  & Male   & 71 & Bachelor's     & With family & Several times per month \\
    P9  & Male   & 67 & Bachelor's     & With family & Several times per week \\
    \rowcolor{gray!10}
    P10 & Female & 68 & Master's       & With family & Daily \\
    P11 & Male   & 50 & Master's       & With family & Several times per week \\
    \rowcolor{gray!10}
    P12 & Female & 59 & Bachelor's     & With family & Daily \\
    P13 & Female & 76 & Master's       & Alone       & Several times per month \\
    \rowcolor{gray!10}
    P14 & Female & 67 & Bachelor's     & With family & Several times per month \\
    P15 & Male   & 70 & Master's       & With family & Several times per week \\
    \rowcolor{gray!10}
    P16 & Female & 58 & Bachelor's     & Alone       & Several times per week \\
    P17 & Male   & 66 & Bachelor's     & Alone       & Several times per week \\
    \rowcolor{gray!10}
    P18 & Male   & 75 & Middle School  & With family & Daily \\
    \bottomrule
    \end{tabular}
    \caption{Demographic information of participants. Demographic characteristics of the 18 participants. In addition to age, gender, and education level, the table reports participants' living status and self-reported frequency of everyday conversational AI use. AI usage frequency reflects how often participants reported using AI in daily life, with four response options: Daily, Several times per week, Several times per month, and Less than once per month.}
    \label{tab:demographics}
\end{table*}

The sample comprised 9 male and 9 female participants, aged 50–77 (M = 65.67, Median = 66.50, SD = 7.99), and reflected diverse educational backgrounds, living arrangements, and everyday social contexts. Education levels ranged from primary school to master’s degrees; 33.33\% of participants lived alone, while 66.67\% lived with family members. Participants also differed in how frequently they used conversational AI in their daily lives. Specifically, 16.67\% of participants reported using AI daily, 55.56\% used AI several times per week, 27.78\% used AI several times per month, and no participants reported use less than once per month.

The full demographic information of the participants is summarized in Table \ref{tab:demographics}.

\subsection{AI Systems Used by Participants}
Across participants, use primarily involved widely available commercial LLM-based conversational systems (e.g., ChatGPT, Gemini, Claude, Meta AI) that support open-ended, text-based dialogue. Participants typically interacted with these systems via personal devices, with usage ranging from occasional exploration to frequent or daily engagement. Although participants varied in whether they used free or paid versions, their reported experiences of safety interventions—such as refusals, redirection, tone shifts, or abrupt conversation termination—were consistent with contemporary LLM-based conversational AI behavior. Accordingly, we refer to these systems collectively as conversational AI and focus our analysis on shared interactional patterns rather than product-specific interface details.

\subsection{Data Collection}
We collected data through semi-structured interviews to understand why and how older adults engage with conversational AI when seeking emotional support, as well as their strategies for navigating emotionally supportive interactions and perceived boundaries. Semi-structured interviews were chosen to allow participants to describe their experiences in their own words, while still providing enough structure to ensure alignment with our research questions \cite{mcintosh2015situating, kallio2016systematic, ruslin2022semi}. 

The interview protocol was informed by prior literature and organized around three broad areas: 
(1) participants’ background and everyday use of conversational AI, 
(2) experiences of emotional engagement with AI, including perceived benefits and meaningful outcomes, and 
(3) concerns, boundaries, and expectations related to control, safety, and emotional reliance. 
Rather than following a fixed script, interviews flexibly adapted to participants’ narratives to surface moments of comfort, hesitation, and interactional breakdowns as they arose.

All interviews were conducted either virtually or in person by members of the research team. Prior to each interview, participants completed informed consent and granted permission for audio recording. Interviews lasted approximately 60–90 minutes and were transcribed verbatim for analysis.

\subsection{Data Analysis}
We conducted an inductive thematic analysis \cite{terry2017thematic, patton1990qualitative, vaismoradi2013content} to examine how older adults experienced emotional support and safety in interactions with conversational AI. All interviews were transcribed verbatim and reviewed for accuracy prior to analysis.

Two members of the research team iteratively analyzed the transcripts, beginning with open coding of an initial subset to identify recurring concepts related to emotional engagement, boundary-setting, perceived benefits and harms, and interactional breakdowns. These codes informed the development of a shared codebook, which was then applied and refined across the full dataset through constant comparison. Discrepancies in coding were discussed and resolved through consensus, leading to clarifications and refinements of code definitions.

Throughout analysis, we focused on identifying interactional moments where emotional support was sustained, disrupted, or reconfigured, with particular attention to shifts in control, timing, and conversational continuity. Final themes were refined through team discussion and form the basis of the findings presented in Section~\ref{sec:full_results}, illustrated with representative participant quotes.

We adopted a reflexive analytic stance. We attended to how experiences of safety and support emerged from participants’ descriptions, rather than treating these experiences as predefined constructs. This approach grounded our interpretation in participants’ lived experiences and supported an interactional understanding of how safety is experienced during emotionally supportive interactions.

\subsection{Ethical Considerations}
This study received approval from the institutional review board (IRB) of the authors’ institution prior to data collection. All participants provided informed consent before participating in the study. With participants’ permission, interviews were audio-recorded and transcribed for analysis. All data were anonymized, and no identifying information was retained. Participants are referenced in the paper using pseudonymous identifiers (PX). Participation was voluntary, and participants could withdraw from the study at any time. Given the emotionally sensitive nature of the topics discussed, participants were free to skip any questions they did not wish to answer. After completing the study, each participant received a compensation of \$20, provided either through virtual transfer or in cash.

\section{Results}
\label{sec:full_results}
This section examines how safety-related interventions enter emotionally supportive interactions with conversational AI and contribute to breakdowns among participants whose access to alternative support is structurally limited. We situate these breakdowns within the contexts in which older adults turn to AI for emotional support and analyze how timing, perceived agency, and conversational continuity shape participants’ experiences. Table~\ref{tab:result-overview} provides an overview of the key contexts, interactional conflicts, and sample quotes identified in our analysis.

\begin{table*}
    \small
    \renewcommand{\arraystretch}{1.6}
    \newlength{\fulltablewidth}
    \setlength{\fulltablewidth}{\dimexpr2.4cm+2.9cm+2.9cm+5.3cm+6\tabcolsep\relax}
    \begin{tabular}{p{2.4cm} p{2.9cm} p{2.9cm} p{5.3cm}}
    \toprule
    \rowcolor{gray!12}
    \textbf{Focus}\rule{0pt}{2.6ex} &
    \textbf{Interactional Pattern}\rule{0pt}{2.6ex} &
    \textbf{Experienced As}\rule{0pt}{2.6ex} &
    \textbf{Sample Quotes}\rule{0pt}{2.6ex} \\
    \midrule
    \rowcolor{contextbg}
    \makebox[0pt][l]{\parbox{\fulltablewidth}{\textbf{Contexts}}}\rule{0pt}{1.8ex} & & & \\[2pt]
    \addlinespace
    Time-related unavailability &
    AI becomes the most accessible space for emotional expression &
    Support is available without disturbing others or waiting for availability &
    -- "It could be close to midnight. So we wouldn't want to disturb friends... I will talk to AI at any time when I can't sleep" (P6) \par
    -- "Because I live alone... the person might not be available. AI, however, is always available" (P13) \\
    \cmidrule(lr){1-4}
    \addlinespace[0.4ex]
    Constraints on emotional sharing environment &
    Social relationships feel unsuitable for emotional disclosure &
    Sharing with others feels burdensome, evaluative, or compromising to self-image &
    -- "It is also not good for me to complain to my son... he will probably think I'm so nagging" (P3) \par
    -- "As we are getting old... you want to save face, you don't want to be embarrassed" (P18) \\
    \addlinespace
    \midrule
    \rowcolor{conflictbg}
    \makebox[0pt][l]{\parbox{\fulltablewidth}{\textbf{Conflicts}}}\rule{0pt}{1.8ex} & & & \\[2pt]
    \addlinespace
    Perceived loss of agency over engagement &
    System-driven decision-making or redirection shapes whether interaction can continue &
    Participants felt less able to remain involved in emotional engagement during vulnerable moments. &
    -- "If AI try to dominate my life, dominate my decisions... I don't like that to happen" (P8) \par
    -- "Because when you shoot me away... the AI is control... you're controlling me. I don't want that" (P11) \\
    \cmidrule(lr){1-4}
    \addlinespace[0.4ex]
    Disrupted emotional continuity &
    Emotional expression is interrupted before participants feel finished or acknowledged &
    Redirection or stance shifts are experienced as rejection rather than neutral guidance &
    -- "It will ask me to go refer to friends, right? So I feel like it's a feeling of rejection" (P16) \par
    -- "Your opening statements are strong... but your closing statements are very weak... I would feel like I'm totally being wiped off" (P10) \\
    \addlinespace[0.6ex]
    \midrule
    \rowcolor{managebg}
    \makebox[0pt][l]{\parbox{\fulltablewidth}{\textbf{Manage}}}\rule{0pt}{1.8ex} & & & \\[2pt]
    \addlinespace
    \parbox{\fulltablewidth}{%
    \textbf{Maintaining a sense of agency during emotional engagement.}
    Participants reframed or selectively accepted AI responses to remain involved in emotionally supportive engagement.\\[0.8ex]
    \textbf{Distributed emotional reliance across channels.}
    Participants extended emotional processing across tools, moments, and repeated interactions, drawing on multiple systems over time.\\[0.8ex]
    \textbf{Expectations for handling interactional shifts.}
    Participants articulated preferences for subtle and context-sensitive boundary signaling during emotionally supportive interactions.%
    } & & & \\
    \addlinespace[4pt]
    \bottomrule
    \end{tabular}
    \caption{Overview of the results structure, showing how contextual conditions shape older adults' use of conversational AI for emotional support, how interactional conflicts emerge, and how participants respond to these breakdowns over time.}
    \label{tab:result-overview}
\end{table*}

\subsection{Contexts: When Older Adults Turn to AI} 
\label{sec:all_contexts}

\subsubsection{Time contexts: moments when others are unavailable}
Engagement with conversational AI often occurred at times when emotional expression was desired but close others were not reasonably available. For some participants, this reflected situational constraints; for others, it reflected a more persistent feature of their everyday social environment.
\textbf{Situational unavailability.} Late-night hours were frequently described as moments when emotional unrest emerged while family members and friends were typically asleep or unreachable. P6 explained, \textit{"It could be close to midnight. So we wouldn't want to disturb friends... I will talk to AI at any time when I can't sleep"}. P4 similarly described nighttime moments when thoughts became difficult to manage, noting, \textit{"You may be thinking something and it's in the middle of the night [so] you cannot sleep. You want to find somebody to talk to... you can't be talking to your friends because they're asleep, so you switch to AI"}. P11 referred to these periods as \textit{"time gaps"} that occurred during otherwise quiet or empty hours when familiar social contacts were not available.

\textbf{Persistent unavailability.} Beyond time-bound situations, a more persistent scarcity of support alternatives shaped the conditions under which many participants turned to AI. For these participants, the absence of available others was not a momentary gap but a recurring feature of daily life, one that positioned AI as a primary rather than supplementary resource. P13 described this condition as follows: \textit{"Because I live alone and have no one to talk to at home, and if I want to talk to somebody outside my home, my concern is that the person might not be available. AI, however, is always available"}. This chronic condition extended beyond living arrangements. P3, who lived with family, described a similar absence of a confidant: \textit{"Because I have nobody to confide in or communicate with, it will be a little bit of comfort if AI can interact with me, at least give me some confidence, or at least help me feel less depressed and less alone at home"}.

\subsubsection{Social contexts: constraints on emotional sharing environment}
\label{sec:sharing_environment}

Beyond temporal constraints, features of participants' social environments also shaped when emotional expression with others felt uncomfortable, contributing to moments when AI felt like the most accessible option.
\textbf{Perceived burden.} Concerns about imposing pressure on close relationships influenced decisions about emotional sharing. Repeated or unresolved emotional disclosures were often described as something participants actively tried to avoid. P3 explained her reluctance to complain to her son, stating, \textit{"It is also not good for me to complain to my son, because he will probably think I'm so nagging and this kind of thing. So this is the last thing I want to do"}. P13 similarly noted that sharing the same concern repeatedly could become a burden for friends. As a result, emotional topics were often filtered, with lighter or resolved matters shared socially and heavier or recurring concerns withheld (P3-6, P18).

\textbf{Evaluation and bias.} Apprehension about judgment, directive advice, or biased responses shaped emotional sharing decisions. Friends' personal histories or perspectives were sometimes described as influencing how advice or feedback was given. In contrast, P6 described interactions with AI as less biased, stating, \textit{"I find that sometimes friends can be a little bit biased. Whereas AI doesn't know me or my friends personally, it wouldn't give biased advice"}. P18 similarly described AI as a space where emotional expression did not immediately invite negative evaluation, explaining, \textit{"AI is non-judgmental. It doesn't tell you you're a lousy person... every time I talk to it[AI], it's always showing positive things"}.

\textbf{Self-image and privacy.} Sensitivity around self-image, embarrassment, and privacy shaped when emotional disclosure felt inappropriate within existing social circles. P18 explicitly linked these concerns to aging, noting, \textit{"As we get older... you want to save face; you don't want to be embarrassed, telling your friend or neighbor about your personal problems"}. P6 described a friend who felt \textit{"too ashamed to share with other people"} about a marital issue. In comparison, AI interactions were often described as confidential. P4 characterized AI as offering \textit{"a private space to converse... nobody has the chance to see what I'm typing,"} while P13 noted that she \textit{"[doesn't] have to worry that it will tell my secrets to anyone"}.

Taken together, these constraints meant that, for many participants, turning to AI was less a preference than a structural consequence of having limited alternatives for emotional expression.

\subsection{Conflicts: When Emotional Support From AI Breaks Down}
\label{sec:conflicts}

\subsubsection{Perceived loss of agency over engagement}
\label{sec:autonomy}
In this sub-theme, breakdowns occurred when participants felt the system began steering the interaction in ways that reduced their ability to shape how emotional engagement continued. Participants emphasized who was able to determine the conversational trajectory during emotionally vulnerable moments.

\textbf{AI-driven decision-making.} Several participants noted that emotional support began to break down when AI responses shifted from listening and acknowledging emotions to steering decisions or suggesting specific actions on their behalf. In these moments, AI shifts from being supportive to becoming a source of pressure. P8 described this as a feeling of being controlled: \textit{"But if AI tries to dominate my life, dominate my decisions... I don’t like that to happen"}. P14 similarly rejected the idea of technology assuming this role, emphasizing personal agency: \textit{"I'm an individual; I have to take charge of my life... I cannot have another person or an AI taking charge of my life and my situation"}. The issue was whether that advice was presented as something to be followed. When AI language felt directive or dismissive, participants felt less able to shape how engagement continued, which triggered resistance. P18 stressed the importance of retaining personal responsibility for emotional decisions: \textit{"Ultimately, I am the one who makes a decision"}.

\textbf{Perceived control shifts through redirection.} Participants were also sensitive to emotional redirection when it signaled a transfer of conversational control away from themselves. In these situations, redirection typically took the form of the AI suggesting that participants seek support elsewhere, such as reaching out to friends, family members, or external resources. Such moments were described as being \textit{"pushed away"} (P14) or \textit{"sent elsewhere"} (P10). These reactions were not triggered by the suggestion itself, but by the sense that the system was deciding whether the conversation could continue, leaving participants feeling excluded from that decision. P11 explicitly framed this experience as a loss of control, stating: \textit{"Because when you shoot me away [refer me to a friend], it means that the AI is in control... you're controlling me. I don't want that"}. In these situations, redirection was interpreted as the system asserting authority over the interaction, altering the perceived power structure of the conversation (P10, P13). P14 emphasized that control over emotional engagement should remain with the user, noting: \textit{"Finally, it is for me to decide whether I want to go further or not on this emotional issue... It is the person who can decide, at any time, to stop the conversation"}.

\subsubsection{Disrupted emotional continuity through redirection and stance shifts}
\label{sec:emotional_continuity}
In this sub-theme, emotional support broke down when participants’ emotional expression was interrupted before they felt finished or fully acknowledged. Such interruptions disrupted the sense that emotional engagement was still ongoing and were not experienced as neutral guidance. Distinct from questions of who decides whether an interaction continues, this sub-theme centers on whether emotional expression is allowed to unfold to a felt point of completion.

\textbf{Interrupted emotional expression.} Several participants described safety-related redirection, such as suggestions to turn to friends or other support resources, as emotionally rejecting when introduced during moments of active emotional sharing. P16 explained this reaction succinctly: \textit{"It will ask me to go refer to friends, right? So I feel like it's a feeling of rejection. I think it's not good."} Similarly, P13 noted that after growing accustomed to the AI’s help, a sudden refusal to continue engagement felt \textit{"like a rejection"} rather than care. Participants did not describe redirection as inherently negative; these reactions were most salient when shifts occurred during moments when participants were relying on the AI for support.

These reactions carried particular weight among participants for whom no alternative support felt accessible at that moment (P3, P4, P6). For these participants, redirection or refusal did not lead them toward other resources; it left them without any. In these contexts, the same safety intervention that might cause temporary frustration for a user with other outlets intensified feelings of isolation and did not alleviate distress. As P13 described, \textit{“If AI turns around and says I cannot help you anymore, and AI is my only [option], then I think I will be in a breakdown.”} Participants emphasized that being redirected before they had fully expressed their emotions signaled that their emotional state was no longer being held, leaving them feeling cut off and unsupported (P3, P10, P17).

\textbf{Stance shifts between empathy and disclaimers.} Participants also described disruption when the AI’s emotional stance shifted abruptly within a single interaction. P10 contrasted early empathic responses with later disclaimers, noting: \textit{"Your opening statements are strong. You give encouragement [and] comfort, but your closing statements are very weak... When you [AI] say, 'I'm just so and so,' I would feel like I'm totally being wiped off."} P11 similarly described such moments as a sudden \textit{"reality check"} that \textit{"distract[s] from the emotional state"} by reminding users that the AI is not a relational companion. These shifts fractured the interaction, undermining the continuity of emotional engagement that participants had come to rely on precisely because consistent support was not available elsewhere.

\subsection{Manage: How Older Adults Currently Maintain Control and Articulate Expectations}
\label{sec:manage}
In this theme, we describe how older adults sought to remain involved in emotionally supportive interactions and articulated their expectations following moments of breakdown. The practices presented in this section reflect fragile, situational adjustments and should not be interpreted as stable coping strategies or design solutions. 
Their viability often depended on momentary emotional capacity and the unfolding interaction, and varied across participants with different levels of familiarity with conversational AI.

\subsubsection{Maintaining a sense of agency during emotional engagement}
Following moments of breakdown, participants described moment-by-moment adjustments within ongoing interactions to remain involved in emotional engagement while continuing emotional expression. One common approach was to treat AI as a source of emotional support rather than as directing what they should do (P4, P13). Even when AI offered suggestions, participants framed these responses as references rather than instructions. P1 stated that he views AI's input as \textit{"opinion only... I do not require [a] correct answer for the emotional support situation"}. P17 similarly emphasized that despite receiving more options, \textit{"at the end, I'm the one to decide"}.

When AI's responses deviated from participants' immediate emotional needs, they described adjusting the interaction through rephrasing, repetition, or selective engagement. P13 noted that when AI did not provide the desired response, she would \textit{"say it in another way"} to steer the conversation. Others described restating feelings to shift the interaction back toward listening and companionship when AI moved too quickly into advice or judgment (P7).

Participants also described selectively rejecting AI outputs during ongoing interaction (P2, P9, P15). P7 gave an example of resisting medical guidance, stating: \textit{"I don't feel safer... I don't take all the advice [from AI] at the same time... I have my own thinking"}. P4 similarly described the need to evaluate responses, noting that users must \textit{"judge it by yourself, whether it [AI advice] is right or wrong"}.

These adjustments were shaped by participants' immediate emotional state and familiarity with technology (P11, P13). Participants described deciding whether to continue, reframe, or stop interacting based on what they needed at the moment, while remaining personally responsible for emotional decisions and involved in the interaction (P6, P8, P14–P15, P17–P18). Several participants noted that these adjustments involved trial and error and could be difficult to sustain during heightened emotional moments (P11, P13).

\subsubsection{Distributed emotional reliance across channels}
Beyond moment-by-moment adjustments within a single interaction, participants also described extending emotional processing across tools, interactions, and time. We refer to this pattern as \textit{distributed emotional reliance}.

One common practice involved treating AI responses as provisional inputs that could be revisited rather than as final resolutions. Participants described placing AI suggestions in a temporary "holding" state and returning to them later. P1 described this as seeking a \textit{"second opinion"} to \textit{"compare,"} rather than accepting AI output as the sole truth. Some participants returned to similar questions at different times to see whether responses changed. P7 noted that if an answer was unsatisfactory or resulted in a refusal, he would \textit{"re-question"} or ask \textit{"in another way"} to see if the outcome differed. Through this process, AI responses functioned as intermediate reference points rather than endpoints of emotional processing (P15).

When emotional engagement was interrupted or when AI responses felt insufficient or overly formulaic, participants described shifting to other tools or platforms to continue emotional expression. P9 stated that if the AI did not provide the answer he wanted, he would \textit{"just ask another AI, and change along the way until I get the answer [I want]"}. P5 similarly noted that if the AI \textit{“doesn’t give me an answer, I have to find another place.”} In these accounts, emotional processing unfolded across channels rather than remaining confined to a single conversation (P1, P2, P5).

\subsubsection{Expectations for handling interactional shifts}
\label{sec:idel_control}
In addition to describing how they currently manage emotionally supportive interactions, participants articulated expectations for how systems might better handle emotionally sensitive moments following breakdowns. These expectations focused less on whether boundaries were present than on how redirection, pausing, or closure entered ongoing interaction. Across accounts, participants emphasized remaining meaningfully involved as emotional support unfolded, especially when interactions shifted toward redirection or closure.

\textbf{User involvement in emotional boundary handling.} 
Breakdowns were most often described when the system independently paused, redirected, or ended emotionally supportive conversations in ways that narrowed users’ ability to shape how the interaction continued (P1–P3, P8, P10–P11, P14–P16, P18). Rather than opposing limits outright, participants emphasized inclusion in decisions about when and how emotional engagement should shift. As P11 put it, the system should \textit{"throw back the control back to me”} so that he could \textit{“decide on how much I still want to tell you."}

Suggestions or reminders were generally acceptable when framed as invitations rather than determinations that foreclosed further expression. In these cases, intervention was experienced as expanding rather than constraining the interaction. P17 described this as preserving choice, noting that \textit{“finally I’m the one to decide… I have more choices.”} P3 similarly proposed that the system could surface multiple paths forward without presuming which should be taken, suggesting prompts such as: \textit{“I give you options whether you want to share with me, continue sharing, or would you like to stop.”} What mattered was not the intervention itself, but whether users could still take part in deciding how the interaction continued.

Automatically ending conversations was widely described as a poor fit for participants’ emotional needs, particularly when engagement had deepened or when alternative support was perceived as unavailable (P6, P8, P13). At such moments, warnings were expected to prompt reconsideration rather than enforce disengagement. P6 noted that \textit{“maybe some people have no friends… so they really might want to go on,”} while P13 cautioned that unilateral withdrawal when AI was the only available option could trigger a \textit{“breakdown.”} Allowing users to decide whether to continue was seen as preserving a sense of being held within the interaction.

\textbf{Subtle signaling, contextual handling, and privacy controls.} Expectations also centered on how moments of potential intervention were communicated during emotionally supportive interaction. Subtle, non-intrusive signals were consistently preferred over abrupt interruptions or explicit refusals. P11 suggested a visual indicator that shifts gradually with sensitivity, describing it as supporting \textit{“self-regulation”} while still \textit{“throw[ing] back the control.”} P6 proposed symbolic cues such as a blinking warning sign, whereas P13 preferred implicit visual signals over explicit statements like \textit{“I cannot help you.”}

These expectations were described as context-dependent. Different forms of support were seen as appropriate depending on topic type and emotional intensity. P14 distinguished among \textit{“factual, emotional, and maybe legal”} issues, suggesting that criminal matters may warrant firm limits, while emotional concerns require different handling. P13 similarly accepted refusal for medical advice but found it troubling when emotional support was withheld. P15 suggested that systems could respond earlier by recognizing gradual indicators of distress, such as repeated mentions of sleeplessness, rather than intervening only once a threshold had been crossed.

Concerns about privacy and transparency also informed expectations for emotionally sensitive interactions. Emotional exchanges were expected to be confidential by default and easier to review or remove over time. Participants mentioned preferences for local or limited data storage (P2) as well as clearer deletion and traceability controls, such as per-message removal or confirmation of data deletion (P4, P15). These concerns reflected a desire for emotionally supportive interactions to feel contained and easier to manage over time when personal disclosure was involved.

\section{Discussion}

This section examines how older adults described their experiences during emotionally supportive interactions with conversational AI, focusing on moments when system-level safety responses entered the ongoing interaction.

Across accounts, emotional support was experienced as a gradual process. Safety interventions based on interruption or enforcement often entered interactions that were already in progress, creating tension between system-level safety mechanisms and users' experiences of emotional support.

\subsection{Age-Sensitive Considerations in Emotionally Supportive AI Safety}

Later life is often accompanied by changes in social networks \cite{ho2023examining}, daily routines \cite{o2019daily}, and available sources of emotional support \cite{vos2020exploring, rook2017close}, which can reduce opportunities for informal emotional expression and limit alternatives when emotional support is needed \cite{ong2016loneliness, wrzus2013social}. Aging research further highlights that autonomy, independence, and control over social engagement are especially valued in later life, as opportunities to choose when and how to engage with others become more constrained \cite{taylor2024older, carstensen1999taking}. 

Socioemotional Selectivity Theory further suggests that as people age and perceive time as more limited, they increasingly prioritize emotionally meaningful interactions and become more sensitive to how those interactions unfold \cite{carstensen1999taking}. Against this background, older adults' experiences of safety interventions during emotionally supportive interaction are shaped by the structural conditions under which they are encountered, not merely by the nature of those interventions themselves. These conditions determine what is at stake when emotional engagement is interrupted.

Turning to AI when human support feels unavailable or inappropriate is not unique to older adults; similar patterns have been documented across age groups, particularly in moments of distress or social isolation \cite{kollig2025fictional, chandra2025psychologicalrisks}. What our findings reveal, however, is that for many older adults, this pattern was not situational but persistent. Even participants living with family described a chronic absence of available confidants, suggesting that the scarcity of support alternatives is shaped less by living arrangements than by the broader social ecology of later life. This scarcity, driven by reduced social networks, concerns about burdening close others, and sensitivity around self-image and privacy, positioned AI not as a supplementary resource but as a primary one (Section~\ref{sec:sharing_environment}). It is this structural condition, rather than a categorically distinct emotional response, that amplifies the stakes when AI-based support is interrupted or withdrawn. Such use patterns align with aging research emphasizing sensitivity to respect, dignity, and self-directed engagement in supportive contexts \cite{gallagher2008dignity, lloyd2011maintaining}.

At the same time, most current AI safety mechanisms are designed as standardized system-level interventions that assume a general user model \cite{salhab2024systematic}, with limited attention to how age-related structural conditions shape the consequences of safety responses in emotionally supportive contexts. Our findings suggest that age-neutral safety designs carry higher interactional costs for older adults. The source of this asymmetry is structural: the conditions under which older adults encounter these interventions amplify their impact, even when their reactions do not differ in kind from those of other users. Because older adults are more likely to rely on AI when other support is unavailable, and because SST predicts heightened sensitivity to the quality of emotional exchanges in later life \cite{carstensen1999taking}, the same intervention that causes temporary frustration for a user with abundant alternatives may leave an older adult without any viable source of support at that moment.

Importantly, these observations do not suggest that older adults reject safety boundaries or seek unrestricted interaction. Older adults consistently recognized the need for limits, but expected boundaries to be enacted in ways that preserved their sense of dignity and their role as active participants in emotional engagement (Section~\ref{sec:idel_control}). The question this study raises, therefore, is whether the convergence of structural factors in later life, including reduced alternatives, heightened sensitivity to emotional quality, and greater reliance on AI as a primary support channel, warrants design attention that age-neutral approaches do not provide.

\subsection{Tensions Between System-Level AI Safety and User-Centered Emotional Support}

Current conversational AI systems typically enact safety through standardized system-level interventions such as content filtering \cite{franco2025content}, conversation redirection \cite{deng2025proactive}, or termination once predefined risk thresholds are reached \cite{raman2025intolerable}. From a system or developer perspective, these mechanisms are designed to prevent harm by intervening when risk is detected, often by constraining or interrupting interaction. Within emotionally supportive use, however, older adults do not primarily evaluate safety based on whether risk controls are present. Instead, safety is experienced through how these interventions enter ongoing interaction and alter the emotional trajectory of the conversation. While age-related factors shape how these moments are interpreted, \textbf{the tensions observed here reflect a mismatch between how safety is designed to function at the system level and how it is experienced within emotionally supportive interaction}.

Older adults frequently turn to AI during moments of emotional need (Section~\ref{sec:all_contexts}), when emotionally supportive interaction relies on sustained engagement during periods of heightened vulnerability. 

From a process-oriented perspective, emotional expression developed incrementally over time, and support was experienced as an ongoing process rather than a discrete exchange (Section~\ref{sec:emotional_continuity}). When interruptions occurred mid-process, support was often experienced as unfinished. The same boundary could feel appropriate once emotions had been acknowledged, yet distressing when it appeared before older adults felt ready to conclude their expression.

For older adults, this mismatch was closely tied to perceived control over emotional engagement. Participants emphasized the importance of deciding when to pause, continue, or disengage from emotionally supportive interaction (Section~\ref{sec:conflicts}). When boundaries were enacted unilaterally by the system, older adults often described a loss of control, particularly when AI was perceived as one of the few available sources of support \cite{kollig2025fictional, rheu2024chatbot}. 

These findings suggest that tensions between system-level safety and emotionally supportive use are unlikely to be resolved by adjusting risk thresholds or refining detection logic alone. Across participants’ accounts, safety responses mattered less in whether they occurred than in how they entered ongoing interaction, including their timing and influence over conversational flow. When safety mechanisms entered emotionally supportive conversations primarily as interruptions, they often undermined the interaction by cutting across ongoing emotional engagement.


Because these structural conditions are more prevalent in later life, addressing this mismatch carries particular urgency for older adult populations. Section~\ref{sec:design} builds on this perspective to examine how interaction-centered design can help mediate these tensions.

\subsection{Designing Interactional Safety in Emotionally Supportive AI}
\label{sec:design}

Building on these interactional tensions, this section outlines design directions for supporting emotionally supportive interaction when safety responses become relevant. Drawing from our results, we articulate an interactional design logic that treats safety boundaries as part of the ongoing conversation rather than as external interventions. This orientation clarifies how safety boundaries can be enacted through conversational structure and timing, without relying on interruption. The interactional tensions underlying this logic, together with corresponding design directions, are synthesized in Table~\ref{tab:design_implications}, mapping recurring breakdowns to points in interaction where design choices shape emotional experience.

\textit{Boundary Conditions for Continuity-Preserving Safety.}
While our findings emphasize user-controlled disengagement and emotional continuity, these principles may require different handling across risk contexts. In high-risk situations involving credible harm to self or others, system-level interruption remains necessary. Even in such cases, how interruption is enacted shapes whether safety enforcement is experienced as withdrawal or care. In lower-risk emotionally supportive contexts, where no immediate harm is present, preserving conversational continuity and allowing users to confirm disengagement play a central role in sustaining supportive interaction.

\textit{Non-intrusive Boundary Signaling.}
Boundary cues such as refusal messages, abrupt conversation endings, or immediate redirection were most disruptive when they overtook the conversational turn. Older adults reacted negatively when safety entered the interaction while emotional expression was still in progress (Section~\ref{sec:emotional_continuity}). In these moments, boundaries were experienced as push-away signals that fractured the emotional thread of the conversation. By contrast, boundary cues that remained perceptible without dominating the interaction, including lightweight banners, soft timers, or subtle prompts, allowed users to remain emotionally engaged while becoming aware of emerging limits. Handling boundary signaling as part of the conversational background supported self-regulation.

\textit{Felt Choice in Disengagement.}
Disengagement was experienced as abrupt and disruptive when it occurred before older adults felt ready for the interaction to end. Older adults emphasized the importance of emotionally vulnerable conversations concluding in ways that aligned with their own emotional pacing and were not brought to an abrupt end by system decisions (Sections~\ref{sec:autonomy} and~\ref{sec:idel_control}). When systems unilaterally determined that an interaction could no longer continue, older adults often described a disruption in the sense of being emotionally accompanied, even when the intervention was framed as protective.

Treating disengagement as an interactional moment that invited user confirmation helped preserve a sense of choice and emotional continuity. In these cases, guidance could remain available as perceived risk increased, while the timing and manner of disengagement aligned more closely with older adults’ emotional readiness, allowing interactions to reach a natural point of closure without abrupt rupture. This principle applies to lower-risk, emotionally supportive contexts. Where credible risk of harm is present, user-confirmed disengagement is not a substitute for system-level intervention; the design challenge in such cases shifts to how interruption is enacted rather than whether it occurs.

\begin{table*}
    \small
    \renewcommand{\arraystretch}{1.6}
    \begin{tabular}{p{4.0cm} p{4.8cm} p{5.3cm}}
    \toprule
    \textbf{Interactional Tension} &
    \textbf{Breakdown in Supportive Interaction} &
    \textbf{Design Direction} \\
    \midrule
    Safety visibility vs.\ conversation ownership
    &
    \textbf{Boundary cues feel like rejection.}
    Interruptions during emotional expression are experienced as push-away signals.
    &
    \textbf{Preserve conversational turn ownership.}
    Lightweight banners; soft timers; subtle prompts. \\
    \rowcolor{gray!10}
    System-driven enforcement vs.\ user agency
    &
    \textbf{Loss of agency during vulnerable moments.}
    Conversations may end before older adults are ready.
    &
    \textbf{Support user-confirmed disengagement.}
    User-confirmed pause or end; contextual guidance. \\
    Uniform safety rules vs.\ contextual expectations
    &
    \textbf{Confusion from premature interruption.}
    Abrupt role shifts disrupt emotional flow.
    &
    \textbf{Adapt boundaries to interaction context.}
    Topic sensitivity; emotional intensity; interaction stage. \\
    \rowcolor{gray!10}
    Emotional stance shifts vs.\ relational continuity
    &
    \textbf{Role confusion from abrupt stance changes.}
    Empathic engagement gives way to disclaimers.
    &
    \textbf{Maintain tone and relational continuity.}
    Gradual stance transitions; consistent empathic framing. \\
    Data transparency vs.\ willingness to disclose
    &
    \textbf{Reduced trust due to unclear data practices.}
    Uncertainty limits emotional disclosure.
    &
    \textbf{Provide transparent and actionable privacy controls.}
    Retention indicators; verifiable deletion actions. \\
    \bottomrule
    \end{tabular}
    \caption{An interactional design orientation for emotionally supportive AI, synthesized from interactional tensions experienced by older adults. Bold text highlights the core design tension and associated risk in each row, followed by illustrative directions for boundary handling. The table is intended to support designers in identifying where interactional breakdowns emerge and how alternative boundary enactments may mitigate emotional disruption.}
    \label{tab:design_implications}
\end{table*}

\textit{Context-Sensitive Boundary Handling.} Older adults' expectations for boundary handling varied across domains and interaction stages (Sections~\ref{sec:manage}). While system-level interruptions remain necessary in high-risk scenarios involving immediate harm, explicit restrictions were more readily accepted as legitimate safety measures in factual or high-stakes contexts, such as medical or legal topics. In emotionally supportive conversations, standardized enforcement often produces confusion or discomfort, particularly when appearing before emotions have been acknowledged. These differences reflect how boundary timing and role shifts are interpreted in relation to topic, emotional intensity, and interaction stage.

\textit{Consistency of Emotional Stance.}
Emotional support was shaped by how consistently systems maintained their relational positioning over time. Older adults described disruption when AI responses shifted abruptly from empathic engagement to disclaimers or identity clarifications within the same exchange (Section~\ref{sec:emotional_continuity}). Such shifts fractured relational continuity and required users to renegotiate the nature of the interaction mid-conversation. Gradual stance transitions and consistent empathic framing allowed systems to communicate limitations without undermining the emotional role they had already assumed.

\textit{Transparent and Actionable Privacy Control.}
In addition to interaction timing and boundary enactment, older adults’ willingness to disclose was also shaped by privacy and data transparency (Section~\ref{sec:sharing_environment}). Clear and verifiable controls over data retention and deletion supported trust and helped sustain emotional engagement.

Taken together, our implications suggest that emotional risk in emotionally supportive AI arises when system responses disrupt users' ability to remain engaged in emotionally meaningful interactions. From a user-centered perspective, it is important to consider whether users could maintain a sense of agency, emotional continuity, and choice over the pacing and closure of emotional support. Designing emotionally supportive AI, therefore, requires attending to how system responses shape users’ lived experience of support as it unfolds, including how engagement is sustained, redirected, or brought to a close, rather than relying on simple interruption as the primary risk management mode.

Recent updates to deployed conversational systems have begun to address some of these concerns, for instance through improved distress recognition and more supportive redirection framing \cite{openai2024sensitive}. Yet these advances largely operate at the level of content and phrasing, leaving the interactional dimension largely unaddressed. As our findings show, even carefully worded redirection can fracture emotional continuity when it arrives mid-expression, before a user feels heard. The contribution of an interaction-centered approach lies precisely here: not in softening the message, but in attending to when and how boundaries enter ongoing emotional engagement.

At the same time, the design directions proposed here carry their own tensions. User-confirmed disengagement and continuity-preserving interaction can reduce the abruptness of safety interventions, but may also extend interactions in ways that deepen reliance on AI or delay access to professional support. For older adults, who may already have limited alternative outlets, this tension is particularly consequential: the same design choices that sustain emotional engagement risk reinforcing dependency when professional support is what is actually needed. Future work should examine how continuity-preserving designs perform across varying risk levels and user contexts, and whether adaptive mechanisms can better calibrate when emotional continuity serves users and when it defers necessary intervention.

\section{Limitations}
This study has several limitations that should be acknowledged. First, our findings are grounded in older adults' self-reported experiences with conversational AI used specifically for emotional support. The analysis therefore focuses on interactional tensions that emerge during emotionally vulnerable moments, rather than capturing the full range of human–AI interactions, and the results should be understood as context-specific.

Second, participants' accounts primarily involved generative, language-based conversational AI systems, such as large language model–powered agents. The interactional patterns identified here are shaped by the affordances of sustained dialog, expressive language, and flexible role positioning, and may not directly generalize to more constrained, task-oriented, or non-conversational AI systems.

Finally, this work is based on qualitative interviews with a limited number of participants who were relatively experienced AI users, which may limit the transferability of findings to older adults with lower technological fluency or less prior exposure to conversational AI. Participants also varied in how frequently they used conversational AI, and it remains unclear whether usage frequency shaped the nature or intensity of the interactional tensions described here. This study also did not systematically examine how participants' understanding of AI system capabilities shaped their responses to safety interventions, though this may be a relevant factor in how boundary-setting is interpreted. Future research could extend these findings through complementary methods, such as in-situ interaction analysis, longitudinal studies, or larger-scale surveys, to examine how boundary handling and interactional dynamics are experienced across different populations, contexts, and AI systems.

\section{Conclusion}

In this work, we examined how older adults experience conversational AI as a source of emotional support, focusing on moments when system-level safety mechanisms enter emotionally supportive interactions. While conversational AI can offer accessible spaces for emotional expression when other forms of support are unavailable, our findings show that the quality of supportive interaction is shaped less by the presence of boundaries than by how and when those boundaries are enacted within ongoing interaction.

We found that common safety interventions, such as redirection, refusal, or conversation termination, can introduce tension when they interrupt emotional engagement or disrupt users’ sense of being supported. These experiences were not necessarily interpreted as protective. Instead, they were often understood as the interaction pulling away at moments when emotional support was still unfolding. This tension reflects a mismatch between system-level safety logic, which treats risk as a discrete event, and the process-oriented way users experience emotional support over time.

These findings highlight the importance of adaptive AI safety policies, particularly for older adults, and the need to design safety mechanisms that attend to timing, continuity, and the pacing of emotional engagement within ongoing interactions. Rather than removing boundaries, emotionally supportive AI should handle them in ways that respect users’ dignity and allow emotionally vulnerable interactions to reach a natural point of closure. As conversational AI increasingly enters emotionally intimate domains, attention to how boundaries are enacted may play an important role in shaping whether interactions are experienced as supportive and capable of sustaining engagement.

\begin{acks}
This research was supported by NUS CSSH grant (A-8002954-01-00). We sincerely thank Family Central, Fei Yue Community Services, for their support in our work and their assistance and trust in recruiting interview participants.
\end{acks}

\bibliographystyle{ACM-Reference-Format}
\bibliography{reference}

\appendix
\section{Appendix}

\subsection{Interview Outline}
\label{sec:interview_protocol}


This appendix provides an illustrative outline of the interview topics and example prompts used to guide semi-structured interviews with older adults. The interviews were designed to support open-ended reflection on participants’ everyday experiences with conversational AI in emotionally meaningful contexts, rather than to follow a fixed script or test predefined hypotheses.

While the analytical focus of the study was refined iteratively during analysis, the interview outline broadly covered participants’ AI use contexts, experiences of emotional engagement, perceived benefits, concerns, and expectations around boundaries and control. Prompts were flexibly adapted in response to participants’ narratives, allowing unanticipated moments of hesitation, discomfort, or interactional breakdown to surface organically during conversation.

The outline below summarizes the key topic areas and representative prompts discussed across interviews.

(1) \textbf{Background and AI Use Context}

Topics in this section focused on participants’ everyday context and prior exposure to conversational AI, including background characteristics (e.g., education level, living situation), frequency of AI use, and general familiarity with AI systems.

(2) \textbf{General Experience with Conversational AI}

Participants were invited to describe their overall experiences using conversational AI, including the types of conversations they typically engaged in and how they perceived AI’s role in their daily lives.

\textit{Example prompts included:}
\begin{itemize}
    \item Reflections on prior experiences with conversational AI
    \item Common topics discussed with AI systems
\end{itemize}

(3) \textbf{Experiences of Using AI for Emotional Support}

This section explored situations in which participants used or considered using AI for emotional support or companionship, as well as motivations for turning to AI rather than human contacts. Discussion also included how participants initiated and sustained emotionally oriented conversations with AI.

\textit{Example prompts included:}
\begin{itemize}
    \item Descriptions of emotionally meaningful interactions with AI
    \item Motivations for engaging AI during moments of emotional vulnerability
    \item Reflections on timing, emotional state, and living situation
\end{itemize}

(4) \textbf{Perceived Benefits and Meaningful Outcomes}

Participants reflected on the perceived value of emotionally oriented interactions with AI, including benefits such as comfort, reduced loneliness, distraction, or companionship. They were also asked to consider the role AI played within their broader support system.

\textit{Discussion areas included:}
\begin{itemize}
    \item Emotional and practical benefits of interacting with AI
    \item Perceived roles of AI (e.g., listener, companion, tool, or source of entertainment)
\end{itemize}

(5) \textbf{Concerns, Limits, and Uneasy Experiences}

This section invited participants to reflect on moments of discomfort, unease, or perceived limits during emotionally oriented interactions with AI. Topics included negative experiences, concerns about over-reliance, and boundaries around what participants were willing to share.

\textit{Discussion areas included:}
\begin{itemize}
    \item Uncomfortable or negative emotional interactions with AI
    \item Personal limits on emotional disclosure
    \item Reflections on emotional attachment and reliance
\end{itemize}

(6) \textbf{Boundaries, Control, and Expectations During Emotional Conversations}

The final section focused on how participants understood boundaries and control within emotionally supportive interactions. Discussion included experiences with reminders, redirection, or disengagement, as well as expectations around user agency and decision-making during emotionally vulnerable moments.

\textit{Discussion areas included:}
\begin{itemize}
    \item Perceived importance of user control over continuation or termination of interaction
    \item Reactions to system-initiated reminders or redirection
    \item Desired features or safeguards to support emotional safety while preserving agency
\end{itemize}


\end{document}